\documentclass[a4paper, 12pt]{article}
\usepackage{amsmath,amsthm,amssymb}
\usepackage{graphicx}
\usepackage{color}
\usepackage{dsfont}
\usepackage{epstopdf}
\usepackage[hang]{subfigure}
\usepackage{natbib}
\usepackage{booktabs}
\usepackage{textcomp}

\usepackage{ulem}
\usepackage{color}

\newtheorem{theorem}{Theorem}

\newtheorem{condition}{Condition}

\definecolor{red}{RGB}{139,0,18}
\definecolor{lightred}{RGB}{186,25,31}
\definecolor{blue}{RGB}{0, 0, 255}
\definecolor{lightblue}{RGB}{69,100,139}
\renewcommand\emph[1]{{\color{red}\itshape #1}}

\newcounter{rmk}
\newcommand\rmk[1]{\vspace*{1mm} \par \stepcounter{rmk}{\noindent \bf Remark \thermk}. {#1}\vspace*{1mm}}

\newcommand\argmin{\mathop{\arg\min}}
\newcommand\tr{\text{tr}}

\newcommand\lam{\lambda}
\newcommand\diag{\text{diag}}
\newcommand\mR{\mathds{R}}
\newcommand\mN{\mathcal{N}}
\def\t{{ \mathrm{\scriptscriptstyle T} }}

\newcommand\sign[1]{\textnormal{sign}{(#1)}}

\newcommand{\normm}[1]{{\vert\kern-0.25ex\vert\kern-0.25ex\vert #1
		\vert\kern-0.25ex\vert\kern-0.25ex\vert}}

\title{Interaction Pursuit Biconvex Optimization}
\author{Yuehan Yang\thanks{School of Statistics and Mathematics, Central University of Finance and Economics},
~~ Siwei Xia$^\dag$~~and~~Hu Yang\thanks{College of Mathematics and Statistics, Chongqing University}
}
\date{}

\begin{document}
\maketitle
\begin{abstract}
Multivariate regression models are widely used in various fields such as biology and finance. In this paper, we focus on two key challenges: (a) When should we favor a multivariate model over a series of univariate models; (b) If the numbers of responses and predictors are allowed to greatly exceed the sample size, how to reduce the computational cost and provide precise estimation. The proposed method, Interaction Pursuit Biconvex Optimization (IPBO), explores the regression relationship allowing the predictors and responses derived from different multivariate normal distributions with general covariance matrices. In practice, the correlation structures within are complex and interact on each other based on the regression function. The proposed method solves this problem by building a structured sparsity penalty to encourages the shared structure between the network and the regression coefficients. We prove theoretical results under interpretable conditions, and provide an efficient algorithm to compute the estimator. Simulation studies and real data examples compare the proposed method with several existing methods, indicating that IPBO works well.
% achieves the lowest bias and MSE.
\end{abstract}

\vspace*{4mm}
\noindent {\bf Keywords:} Graphical structure, High-dimensional data, Laplacian smoothness, Multivariate regression, Lasso.

\section{Introduction}
Many large-scale statistical applications involve building interpretable models, linking a large set of predictors to a number of responses, such as protein-DNA associations \citep{zamdborg2009discovery}, brain activity predictions \citep{liu2015calibrated} and stock market associations \citep{liao2008mining}. Multivariate regression models have been applied to this kind of task, to find which features are important for determining the responses and to capture the complex structures within the responses. In practice, if we allow both numbers of predictors and responses larger than the number of observations, responses always depend on small fractions of predictors. The correlations between responses could be affected by the inner structures of the predictors and that of the noise matrix, as well as the overlap of fractions of predictors in regression function, making it hard to achieve a good estimate.

Reviewing the literature, since the conditional dependence can capture the direct link between two variables when other variables are conditioned on, many researchers introduce the Gaussian graphical models and estimate the corresponding precision matrix to explore the relationship within variables \citep{yuan2007model,cai2011constrained}. To identify the multivariate regression models, some literature uses a joint regularization penalty for both the regression coefficients and the precision matrix of noises, and solve them iteratively, rendering the heavy computational cost \citep{yin2011sparse,rothman2010multivariate}. \citet{marchetti2019penalized} proposed to regress the eQTL mapping and genes incorporating a Gaussian graphical model over the latter. The defined inverse-covariance-fused lasso procedure is difficult to solve for a large set of variables too, for the fused penalty leads to a quadratic programming problem \citep{tibshirani2007spatial}. Some literature considers an uncorrelation structure and calibrates regularization for each regression with respect to its noise level \citep{liu2015calibrated}. The reduced-rank regression is another effective approach in the multivariate models where the dimension reduction is achieved by constraining the coefficient matrix to have low-rank \citep{bunea2007sparsity,chen2012sparse,mukherjee2015degrees,chen2013reduced,li2017parsimonious}. The structure among multi-response interaction models is also discussed by \citet{cai2013covariate,cai2016estimating}, \citet{molstad2016indirect} and \citet{zou2017covariance}.

Although recent work dedicated to the multivariate models, there's still a gap of our understanding that most of the previous research either focused on the graph estimation of noises \citep{yin2011sparse,rothman2010multivariate,cai2013covariate,cai2016estimating} or the model is assumed to have some specific format, i.e. block-structured regularization \citep{obozinski2011support}, uncorrelation structure for the noise matrix \citep{liu2015calibrated}. Yet, in many real applications of multivariate models, the structure within responses is not only affected by the correlation of the noises, but also affected by the regression function acting on the predictors, and the correlation within predictors as well.

This paper considers a study where allows both predictors and responses to have their complex structures. To be specific, responses are distributed based on the distributions of noise terms, predictors and the coefficient matrix, furthermore, each response is allowed to link with different fractions of predictors. Our aim is to identify the correct model under this complex environment. The interactions in both predictors and responses affect each other directly hence the information of the related precision matrices is useful and should not be ignored. Also, the computational cost should be concerned. We present a new procedure, called Interaction Pursuit Biconvex Optimization (IPBO), to address these needs. A key characteristic of IPBO sorting out the above problems is to use the Laplacian quadratic associated with the graph information to promote smoothness among coefficients associated with the correlated predictors and the correlated responses.

Another characteristic of IPBO is that this method explores the graph information directly from the responses and predictors, not from the noises. The reason is twofold: First, the knowledge of noises is hard to figure out in practice. An iterative algorithm may approximate the estimator \citep{rothman2010multivariate}, still resulting in computational burden. Second, the graph estimation of the noises doesn't capture any patterns sharing with the regression matrix, since it only encodes the structure in responses that cannot be explained by predictors \citep{marchetti2019penalized}. We shall discuss the merits of this method and give a more detailed comparison in the next section.

We organize the rest of the paper as follows. Section 2 introduces the method for $ X $ with the uncorrelation structure and the general structure. Section 3 shows the coefficient matrix estimator and related theoretical properties. The simulations and application in Section 4 and Section 5 analyse the performance of IPBO and compare it with several existing methods. We conclude in Section 6. Technique details are provided in the Supplementary Material.

\section{Methods}
% In this section, we present the details of IPBO and compare it with existing methods.
Considering a multivariate regression problem:
\[ Y = X B + E, \]
where $ Y, E \in \mR^{n \times q} $ are matrices of responses and noises, $ X \in \mR^{n \times p} $ is a covariate matrix and $ B = (\beta_{jk})_{p \times q } $ is a matrix of regression coefficients.
% with $ B_{\cdot k} $ denotes the $ k $th column of $ B $ and $ B_{j\cdot} $ denotes the transposition of the $ j $th row of $ B $.
We note that the dimension of $ p $ and $ q $ are allowed to greatly exceed the sample size, i.e. they are allowed to grow at an exponential rate in sample size. For notational simplicity, we do not index them with $ n $. Assume $ X $ be the random samples of a multivariate normal distribution $\mN_p(0, \Sigma)$ and $ E $ be the random samples of $ \mN_q(0,\Lambda) $, then we obtain the structure of $ Y $: $ Y $ is the random samples of a multivariate normal distribution that $ Y | X \sim \mN_q(XB,\Lambda) $ and $ Y \sim \mN_q(0,\Theta^{-1} ) $ where $ \Theta^{-1} = \Lambda + B^\t \Sigma B $ and $ \Theta =(\theta_{kk'})_{q \times q}$ is set to be the precision matrix (inverse covariance) of $ Y $.

\subsection{Uncorrelation structure for $ X $}
We first introduce a less complex structure that the predictors are uncorrelated. It can be seen as a special case for the complete version of IPBO showed in the next section. Since we always normalize the predictors, we can simply assume that
\[ X \sim \mN_p(0, I), \]
then based on the multivariate regression model, we have
\[ Y \sim \mN_q(0, \Lambda + B^\t B ).\]
Let $ B_{j\cdot} $ be the transposition of the $ j $th row of $ B $.
Given $ n $ i.i.d.observations of $ X $ and $ Y $, we define the estimator $ \hat B $ and $ \hat \Theta $ by solving following biconvex optimization:
\begin{align}\label{eq first}
(\hat B, \hat \Theta ):=\argmin _{B, \Theta} & \frac1{n}\|Y-X B\|_{F}^2+\frac1{n} \tr(Y^\t Y \Theta)-\log \det(\Theta) \notag\\
&+\lambda_1\|B\|_1+\lambda_2\|\Theta\|_1 +\gamma \sum\limits_{j=1}^p B_{j\cdot}^\t \Gamma B_{j\cdot},
\end{align}
where i) the first term $\|Y-X B\|_{F}^2/n $ is the regression loss; ii) the second term $ \tr(Y^\t Y \Theta)/n$\\$-\log \det(\Theta) $ is the inverse covariance loss which is derived from the marginal log likelihood of $ y $; iii) the third term $ \|B\|_1 $ and the fourth term $ \|\Theta\|_1 $ are $ l_1 $ regularization functions while $ \lam_1 $, $ \lam_2 $ are tuning parameters; iv) the final term $ \gamma \sum_{j=1}^p B_{j\cdot}^\t \Gamma B_{j\cdot} $ is a Laplacian quadratic penalty and $ \Gamma $ is the Laplacian matrix, a symmetric matrix representation of a graph. For $ j = 1,\dots,p $, this penalty satisfies
\[ B_{j\cdot}^\t \Gamma B_{j\cdot} = \sum_{1 \leqslant k < k' \leqslant q} |\hat \theta_{kk'}|(\beta_{jk} - \sign{\hat \theta_{kk'}}\cdot \beta_{jk'})^2, \]
where $ \hat \theta_{kk'} $ is the element of the estimated precision matrix. Above equality holds since the Laplacian matrix $ \Gamma $ is defined by
\[ \Gamma = D - A, \]
where $ A = \{a_{kk'}\}_{q \times q} $ called the adjacency matrix, and $ D = \text{diag}(d_1,...,d_q) $ with $ d_k = \sum_{k'=1}^q |a_{kk'}| $. The proposed method conducts the adjacency matrix by the estimated precision matrix that $ A = \hat \Theta $ to encourage smoothness among the coefficients of the closely related responses.

This function also can be used when $ X $ are assumed to be fixed, which could associate with another method, MRCE \citep{rothman2010multivariate}, applying the graph information from the noise matrix. To show the merits of the proposed procedure, we use MRCE as a contrast:
\begin{align*}
(\hat B, \hat \Theta^0) =  \argmin_{B,\Theta^0} & \dfrac1{n}\tr\Big[ (Y - XB)^\t (Y - XB)\Theta^0 \Big] - \log \det (\Theta^0) + \lam_1 \|\Theta^0\|_1 + \lam_2 \|B\|_1,
\end{align*}
where $ \Theta^0 $ denotes the inverse noise covariance matrix. This estimator has the following two drawbacks for many applications:
\begin{itemize}
\item[] \textbf{Computational issue.} The algorithm solving MRCE must be iterative with $ (Y - X\hat B)^\t (Y - X\hat B)/n $ updated in every step. As criticized by \citet{rothman2010multivariate}, this algorithm may take many iterations to converge for high-dimensional data and the computational cost is heavy. Though the authors proposed an approximate three-stage algorithm, the first step estimates $ \hat B $ is obtained by assuming an uncorrelation structure for noises, which is far from the true model. There's no guarantee that the solution of this approximate algorithm would close to the solution of MRCE.

\item[] \textbf{Lack of information.}
MRCE assumes that the correlation of the response variables arises only from the correlation in the noises. It doesn't consider the effect of the regression function acting on the predictors. Furthermore, this procedure lucks of some structured sparsity penalty to encourage the shared structure between the graph and the regression coefficients.
\end{itemize}

As a solution to above drawbacks, the second term of \eqref{eq first}, $ \tr(Y^\t Y \Theta)/n-\log \det(\Theta) $, considers the graph information directly from the conditional dependencies of responses, which is quite easier to be obtained than that of noises; the final term of \eqref{eq first}, $\!\gamma \! \sum_{j=1}^p \! B_{j\cdot}^\t \Gamma B_{j\cdot} \!$, promotes smoothness among the coefficient estimation associated with the linked responses.

In this section, we learn the information of the conditional correlation structure of responses with the knowledge that it is not only related to the distribution of the noise matrix but also affected by the coefficient matrix and its sparsity structure as well. We will make full use of the shared structure information by using the complete version of IPBO, see Section 2.2 for more details. In the meantime, we will introduce in Section 2.3 that our algorithm is simple and stable.

\subsection{General $ X $}
We now define the complete version of the IPBO estimator for allowing both $ X $ and $ Y $ have high dimensional sparse complex graph structures. More precisely, we allow many predictors and responses are conditionally correlated, both numbers of which are not too much comparing the overall numbers, i.e. less than the sample size. We are interested in how features interact with each other and try to use this graph information to improve the efficiency of the multivariate regression modeling. Assume that
\[ X \sim \mN_p(0, \Sigma), \]
and we have
\[ Y \sim \mN_q(0, \Lambda + B^\t \Sigma B ).\]
Let $ \Omega$ and $ \Theta$ denote the precision matrices of $ X $ and $ Y $ respectively. We use the penalized negative loglikelihood function $ g(\cdot) $ to estimate two matrices, such as,
\begin{align*}
g(\Theta, \lam_2) = \frac1{n} \text{tr}\left(Y^\t Y \Theta\right)-\log \det(\Theta)+\lambda_2\|\Theta\|_1,\\
g(\Omega, \lam_3) = \frac1{n} \text{tr}\left(X^\t X \Omega\right)-\log \det(\Omega)+\lambda_3\|\Omega\|_1.
\end{align*}
Let $ B_{\cdot k} $ be the $ k $th column of $ B $ and $ B_{j\cdot} $ be the transposition of the $ j $th row of $ B $. Given $ n $ i.i.d.observations of $ X $ and $ Y $, we define the estimator $ \hat B $, $ \hat \Omega $ and $ \hat \Theta $ by solving following optimization:
\begin{align}\label{eq final}
(\hat B, \hat \Theta, \hat \Omega):=\argmin _{B, \Theta,\Omega} &\frac1n \|Y-X B\|_F^2+\lambda_1\|B\|_1+g(\Theta, \lam_2)+g(\Omega, \lam_3) \notag\\
&+\gamma_1\sum\limits_{j=1}^p B_{j\cdot}^\t \Gamma_1 B_{j\cdot}+ \gamma_2\sum\limits_{k=1}^q B_{\cdot k}^\t \Gamma_2 B_{\cdot k}.
\end{align}
Comparing to \eqref{eq first}, \eqref{eq final} adds the estimation of the precision matrix of $ X $ and adds another $ l_2 $ penalty $\gamma_2\sum\limits_{k=1}^q B_{\cdot k}^\t \Gamma_2 B_{\cdot k}$ to encourage similarity between the coefficient estimation of the conditional correlated predictors. The Laplacian matrices, $ \Gamma_1 $ and $ \Gamma_2 $, are conducted by the estimated inverse covariance of $ Y $ and $ X $ respectively. Throughout the paper, the penalized likelihood method for estimating the precision matrix in the Gaussian graphical model was proposed by \citet{yuan2007model}. \citet{friedman2008} developed a fast and stable algorithm called graphical Lasso, which solves a $ 1000 $-node problem ($\sim $500,000  parameters) in a minute. Further, \citet{ravikumar2011high} analysed its performance under high-dimensional scaling.

\subsection{Computational Algorithm and Solution}
Set the objective as
\[f(B,\lam_1)=\frac1n \|Y-X B\|_F^2+\lambda_1\|B\|_1.\]
We present a  two-stage approximate algorithm for solving IPBO:
\begin{center}
    \textbf{Two-stage Algorithm}
\end{center}

Step 1: Initialize $ \hat B^{(0)} = 0 $, seek the minimizer $ \hat \Theta $ and $ \hat \Omega$ of
\[\hat \Theta :=\argmin_{\Theta} g(\Theta, \lam_2), \]
\[\hat \Omega :=\argmin_{\Omega} g(\Omega, \lam_3). \]

Step 2: Let $\Gamma_1 = D_1 - \hat \Theta$ and $ \Gamma_2 = D_2 - \hat \Omega $. Among, $ D_1 = \diag(d_{11},...,d_{1q}) $ where $ d_{1k} = \sum^q_{k'=1}|\hat \theta_{kk'}|$ and $ D_2 = \diag(d_{21},...,d_{2p}) $ where $ d_{2j} = \sum^p_{j'=1}|\hat \omega_{jj'}| $. Seek the minimizer $ \hat B $ of
\begin{align}\label{eq beta}
\hat B:=\argmin_{B}  f(B,\lam_1) + \gamma_1\sum\limits_{j=1}^p B_{j\cdot}^\t \Gamma_1 B_{j\cdot} + \gamma_2\sum\limits_{k=1}^q B_{\cdot k}^\t \Gamma_2 B_{\cdot k}.
\end{align}

The first step gives the estimations of both Gaussian graphical models for the predictors and the responses respectively. Unlike the estimation of the precision matrix of noises, the former does not need iteration. The second step, estimating $ \hat B $ given $ \Gamma_1 $ and $ \Gamma_2 $ where the graph information is introduced fully by both Laplacian penalties, does not need iteration too.

As a contrast, we present a iterative algorithm which considers the above $ \hat \Theta $, $ \hat \Omega $ and $ \hat B $ as the initial estimation and adds Step 3 as following:

Step 3: Repeat the following step until convergence: Given $ \hat B^{(m)} $, seek the minimizer $ \hat \Theta^{(m+1)} $, $ \hat \Omega^{(m+1)} $ of
\[\hat \Theta^{(m+1)} :=\argmin_{\Theta} g(\Theta,\lam_2) +\gamma_1\sum\limits_{j=1}^p (\hat B^{(m)}_{j \cdot})^\t \Gamma_1 \hat B^{(m)}_{j \cdot}, \]
\[\hat \Omega^{(m+1)} :=\argmin_{\Omega} g(\Omega,\lam_3)
+\gamma_2\sum\limits_{k=1}^q (\hat B^{(m)}_{\cdot k})^\t \Gamma_2 \hat B^{(m)}_{\cdot k}, \]
and seek the minimizer $ \hat B^{(m+1)} $ of
\[ \hat B^{(m+1)} :=\argmin_{B} f(B,\lam_1)+\gamma_1\sum\limits_{j=1}^p B_{j\cdot}^\t \Gamma^{(m+1)}_1 B_{j\cdot}+ \gamma_2\sum\limits_{k=1}^q B_{\cdot k}^\t \Gamma^{(m+1)}_2 B_{\cdot k}.\]
Based on the theoretical discussion, both $ \gamma_1 $ and $ \gamma_2 $ are restricted to take small values, hence we have found in simulations that the difference between two algorithms is typically small. During simulations and empirical results, we use the two-stage approximate algorithm to solve the estimator.

To go beyond the basics, we turn to write the fit and solutions of this two-stage algorithm of IPBO. Given $ \hat \Omega = (\hat \omega_{jj'})_{p \times p} $ and $ \hat \Theta = (\hat \theta_{kk'})_{q \times q} $, the optimization \eqref{eq beta} is strictly convex and the derivation of Laplacian penalties are
\[\partial\big\{\sum\limits_{j=1}^p \hat B_{j\cdot}^\t \Gamma_1 \hat B_{j\cdot} \big\}/\partial \hat \beta_{jk}  = \hat \beta_{jk} \cdot \sum_{ k' \neq k}|\hat \theta_{kk'}| - \sum_{ k' \neq k} \hat \theta_{kk'}\hat \beta_{jk'} \triangleq H_1, \]
\[\partial\big\{\sum\limits_{k=1}^q \hat B_{\cdot k}^\t \Gamma_2 \hat B_{\cdot k} \big\}/\partial \hat \beta_{jk} = \hat \beta_{jk} \cdot \sum_{ j' \neq j}|\hat \omega_{jj'}| - \sum_{j' \neq j} \hat \omega_{jj'}\hat \beta_{j'k} \triangleq H_2. \]
Let
\[ H_3 =\partial f(\hat B,\lam_1)/\partial \hat \beta_{jk} =- \dfrac{2}{n}X^\t_j Y_k + \dfrac{2}{n}X^\t_jX \hat B_{\cdot k} + \lam_1 \sign{\hat \beta_{jk}}, \]
where $X_j$ and $Y_k$ represent the $j$th and $k$th column of $X$ and $Y$, respectively. According to the Karush-Kuhn-Tucker (KKT) optimality conditions, for $ \hat \beta_{jk} \neq 0 $, we have
\[ H_1 + H_2 + H_3 = 0. \]
More specifically, above equality can be written as
\begin{align}\label{eq cor}
    & \big(\sum_{ k' \neq k}|\hat \theta_{kk'}|+ \sum_{ j' \neq j}|\hat \omega_{jj'}|  + \dfrac{2}{n}X^\t_jX_j \big)\hat \beta_{jk}\nonumber\\
    & = \big[\dfrac{2}{n}X^\t_j (Y_k - \sum_{j'\neq j}X_{j'}\hat \beta_{j'k}) -\lam_1 \sign{\hat \beta_{jk}} \big]+\sum_{j'\neq j}\hat \omega_{jj'} \hat \beta_{j'k} + \sum_{ k' \neq k} \hat \theta_{kk'}\hat \beta_{jk'}.
\end{align}
If the $ j $th predictor is found to be conditional correlated with other predictors, unlike the $ l_2 $ smooth penalty doing the shrinkage equally, the estimated precision matrix produces the conditional correlation differentially and precisely. It directly affects the associated coefficient estimates, as found on the second term of right-hand side of \eqref{eq cor}. The estimated precision matrix of predictors has the same effect on the coefficient estimate, see the last term of right-hand side of \eqref{eq cor}.

\section{Theoretical Result}
Consider the dimensionalities that the number of predictors $p=O(e^{n^{c_1}})$ and the number of responses $ q=O(e^{n^{c_2}}) $ where $0<c_1,c_2<1$. Let $ S_k = \{j =1,...,p~|~ \beta_{jk} \neq 0 \}$ and $ m_k = |S_k| $ for $k=1,\dots,q$. Let $m=\sum_{k=1}^q m_k$ and assume $m = O(n^{c_3}) $ where $0<c_3<1-c_1$. Particularly, let $ S_\Omega = \{ j,j'=1,...,p~|~\omega_{jj'} \neq 0 \} $, $ S_\Theta = \{ k,k'=1,...,q~|~\theta_{kk'} \neq 0\} $ and $S=\{j=1,\dots,p,k=1,\dots,q~|~\beta_{jk}\neq 0 \}$. For all $ j  = 1,\dots,p $, set the number of nonzero entries of $ \Omega_{j} $ are bounded by a maximum degree, i.e.
\[ d := \max_{j=1,\dots,p}\big|\{j'=1,...,p~|~\omega_{jj'} \neq 0\}\big|, \]
and similar definition on $ \Theta $. This is a natural requirement for the Gaussian graphical model \citep{ravikumar2011high}, and we require that the estimated precision matrices have the same maximum degree too. The following condition is needed for establishing the rate of convergence.

\begin{condition}\label{assum1}
Restricted eigenvalue condition: There exists a positive constant $ \tau_{\min} $ that
\begin{align}
\| X\Delta \|^2_2/n \geqslant \tau_{\min}\|\Delta\|^2_2, ~ \text{for all}~\Delta \in B
\end{align}
and $B= \{\Delta \in \mR^p~\text{and}~|A| < n:\left\| \Delta_{A^c} \right\|_1 \leqslant 7\| \Delta_A\| \} $.
\end{condition}

\rmk{Restricted eigenvalue condition requires the lower bound of the eigenvalues of $ X^\t X / n $ associated with the support. This condition is widely used to bound the $ l_2 $-error between $ \beta $ and the estimate in the uni-response regression models \citep{bickel2009simultaneous,meinshausen2009lasso}, and is the only condition for IPBO obtaining the error bound.}

\begin{theorem}\label{thm 1}
Suppose Condition~\ref{assum1} holds. Set $\lam_1 \!=\! K_1\sqrt{\log p/n}$ and assume $ K_2 \max(\gamma_1, \gamma_2) $\\$\max_{j,k}\{|\beta_{jk}|\} \leqslant \lam_1 $ where $ 0<K_1,K_2<\infty $. There exist positive constants $ c < \min\{c_1,c_2\} $ and $ K $ that with probability at least $ 1 - o(e^{-n^c}) $ we have
\[\|\hat B - B\|_F < K \sqrt{m \log p /n}.\]
\end{theorem}
\rmk{$ \gamma_1 $ and $ \gamma_2 $ are restricted to take small values comparing with $ \lam_1 $. That's natural because large $ \gamma_1 $ and $ \gamma_2 $ would make the smooth penalties become prevalent and the estimates will be hard to set to zero.}

\rmk{Since both smoothing penalties add extra uncontrollable elements in the error bound, the theoretical advantages of these penalties may not be very prominent. The error bound of IPBO is roughly comparable with that of the Lasso.}

\noindent The following condition is needed for recovering the true underlying sparse model.
\begin{condition}\label{assum2}
Irrepresentable condition: Let $C_{S_k}=X^\t_{S_k}X_{S_k}/n$ and $C_{S^c_k}=X^\t_{S^c_k}X_{S_k}/n$. There exists a fixed parameter $ \eta \in (0,1)$ such that for $ k = 1,\dots,q $,
\[ \|C_{S^c_k} C^{-1}_{S_k}\|_\infty \leqslant 1 - \eta, \]
% Irrepresentable condition: There exists a fixed parameter $ \eta \in (0,1]$ such that for $ k = 1,\dots,q $,
% \[ \|\Sigma_{S^c_k S_k} \Sigma^{-1}_{S_k S_k}\|_\infty \leqslant 1 - \eta, \]
and for $ \bar T = \Sigma \otimes \Sigma $ and $\widetilde T = \Theta^{-1} \otimes \Theta^{-1} $ with $ \otimes $ denoting the Kronecker matrix product,
\[ \big\{\max_{e \in S_\Omega^c}\|\bar T_{e，S_\Omega}(\bar T_{S_\Omega S_\Omega})^{-1}\|_1\big\} \vee \big\{\max_{e \in S_\Theta^c}\|\widetilde T_{e，S_\Theta}(\widetilde T_{S_\Theta S_\Theta})^{-1}\|_1\big\} \leqslant 1 - \eta.\]
\end{condition}

\rmk{Irrepresentable condition is well known from previous work on the variable selection consistency of the Lasso and the graphical Lasso \citep{zhao2006lasso,ravikumar2011high}. The proposed procedures and the algorithms minimize both the squared error loss and the loglikelihood loss regularized with the $ l_1 $ penalty, hence this condition is needed for the theoretical support.}

Though there are some other selection penalties for penalized likelihood function and penalized linear function which may not require the irrepresentable condition, we found in simulations and empirical experiences that the $ l_1 $ penalty is more stable and often outperforms others in dealing with complex correlated datasets.

\begin{theorem}\label{thm 2}
Suppose Condition~\ref{assum1}-\ref{assum2} hold. If $\lam_1= K_1\sqrt{\log p/n}$, $ K_3\lam_1 \leqslant  \min_{j,k \in S}\{|\beta_{jk}|\}$ and $ K_2 \max(\gamma_1, \gamma_2) \max_{j,k}\{|\beta_{jk}|\} \leqslant \lam_1 $ where $ 0<K_1,K_2,K_3<\infty $, then with $ 0 < c < \min\{c_1,c_2\} $ the following event holds:
\begin{align*}
P(\hat S=S)\geqslant 1 - o(e^{-n^c}) \rightarrow 1, \ as \ n\rightarrow\infty.
\end{align*}
\end{theorem}

\section{Simulations}
In this section, we present the performance of the proposed methods from six simulation examples, comparing with MRCE \citep{rothman2010multivariate}, GFlasso \citep{chen2010graph} and the $l_1/l_2$ regularization \citep{obozinski2010joint,obozinski2011support}. In what follows, we write $ l_{21} $ short for the $l_1/l_2$ regularization. Let SIPBO (Short for Simplified IPBO) be the abbreviation of the proposed method with only one smoothness penalty on the coefficient estimation and one graph estimation on the responses; IPBO be the abbreviation of the complete version of the proposed method including two smoothness penalties on the coefficient estimation and two graph estimations on the responses and predictors respectively.

Within each example, we fix the sample size $n=100$ and vary the dimensionalities that $p=\{100,150,200,250,300,350\}$, $q=\{50,100,150,200,250,300\}$. Generate the predictor matrix $X$ and the noise matrix $E$ with rows drawn independently from $\mN_p(0,\Sigma)$ and $\mN_q(0,\Lambda)$. Specifically, first four examples use the same coefficient matrix which is defined in Example 1:
\begin{description}
    \item Example 1. $\Lambda=I$ and $\Sigma$ is block diagonal with $ 5 \times 5 $ blocks. For each block, the elements equal $0.8^{|d-d'|}$ where $d,d'=1,\dots,5$. For the coefficient matrix, $B=(\beta_{jk})_{p\times q}$, where $\beta_{jk}=3$ when $k=3j-2,\dots,3j+7$ and $j=1,\dots,5$; $\beta_{jk} = 0$ for others.

    \item Example 2. $\Sigma=I$ and $\Lambda$ is block diagonal with $ 5 \times 5 $ blocks. For each block, the elements equal $0.3^{|d-d'|}$, where $d,d'=1,\dots,5$.

    \item Example 3. $\Sigma=I$ and $\Lambda=I+T_1$. For the $ q \times q $ matrix $ T_1 $, the off-diagonal elements of first $22\times 22$ partition matrix equal $0.6^{|d-d'|}$, where $ d,d'=1,\dots,22 $ and $d\neq d'$; other elements equal zero.

    \item Example 4. The elements of $\Lambda$ and $\Sigma$ equal $0.5^{|j-j'|}$ where $j,j'=1,\dots,p$, and $0.3^{|k-k'|}$ where $k,k'=1,\dots,q$, respectively.

    \item Example 5. $\Sigma=I+T_2$ and $\Lambda=I+T_3$, where the first off-diagonal elements of the first $10\times 10$ partition matrix of $T_2$ and $T_3$ equal $0.5$ while the rest elements of both matrices equal zero. For the coefficient matrix $B$, the first $10 \times 10 $ coefficients equal 3 and others equal 0.

    \item Example 6. $\Sigma=I+T_4$ and $\Lambda=I+T_5$, where the first off-diagonal elements of first $20\times 20$ partition matrix of $ T_4 $, and that of the first $30 \times 30$ partition matrix of $T_5$, equal $0.5$. The rest elements of both matrices equal zero. For the coefficient matrix $B$, the first $20 \times 30 $ coefficients equal 3 with probability 0.6 and others equal 0.
\end{description}

We apply the two-stage algorithm \eqref{eq beta} to solve SIPBO and IPBO. For the first stage that estimating the graphs, we use 5-fold cross-validation to choose $\lam_2$ and $\lam_3$. Then, for the second stage with fixed $\hat\Theta$ and $\hat \Omega$, we use the BIC criterion to choose turning parameters $\lam_1$, $\gamma_1$ and $\gamma_2$:
\begin{align*}
    \text{BIC}(\lam_1,\gamma_1,\gamma_2)&=\sum\limits_{k=1}^q \text{BIC}_k(\lam_1,\gamma_1,\gamma_2)
    =\sum\limits_{k=1}^q \Big(n\times \log(\text{RSS}_k(\lam_1,\gamma_1,\gamma_2))+\log n \times |S_k|\Big),
\end{align*}
where $\text{RSS}_k(\lam_1,\gamma_1,\gamma_2)=\|Y_k-XB_k\|_2^2$ and $|S_k|$ is the cardinality of set $S_k$. $\lam_1$, $\gamma_1$ and $\gamma_2$ are selected by minimizing $\text{BIC}(\lam_1,\gamma_1,\gamma_2)$. The average of each measure is presented base on 100 simulations.

As shown in Table~\ref{tablel2}-\ref{tablemse}, we can see that both SIPBO and IPBO nearly uniformly outperform other methods in both $ l_2 $ error and MSE. SIPBO nearly performs best in Example 1 - 3, which construct the correlation structures either in the predictors or in the responses. Besides, Example 1 - 2 construct the structures containing relevant and irrelevant predictors while Example 3 constructs the correlation structure only in the relevant predictors. We are interested in how much interference the estimates may get from the irrelevant predictors which are correlated with the relevant predictors.
As we can see, both SIPBO and IPBO are not affected much as their performances in Example 1 - 3 behave similarly. Example 4 - 6 considers more complex scenarios that both predictors and the responses have correlation structures. Furthermore, Example 6 considers the case that the nonzero coefficients are randomly assigned. In these cases, IPBO nearly performs the best, followed by the SIPBO.

\begin{table}[!htp]
    \centering
    \small
    \setlength\tabcolsep{6pt}
    \caption{Performance comparison measuring by $l_2$ error.}
    \label{tablel2}
    \scalebox{0.8}{
    \begin{tabular}{llccccc}
    \hline\hline
    Method&q/p &SIPBO&IPBO&MRCE&GFlasso&$l_{21}$\\
    \hline
    Example 1	&	50/100	&	3.108	(0.631)	&	3.338	(0.613)	&	4.163	(0.909)	&	10.863	(1.608)	&	7.908	(1.111)\\
	&	100/150	&	3.213	(0.597)	&	3.467	(0.607)	&	4.553	(0.996)	&	13.099	(1.402)	&	13.532	(1.685)\\
	&	150/200	&	3.456	(0.595)	&	3.602	(0.577)	&	5.708	(1.272)	&	20.351	(1.291)	&	17.43	(1.677)\\
	&	200/250	&	3.443	(0.519)	&	3.644	(0.615)	&	5.729	(1.261)	&	21.067	(2.462)	&	21.81	(2.242)\\
	&	250/300	&	3.498	(0.636)	&	3.826	(0.609)	&	5.875	(1.273)	&	24.101     (1.243)	&	25.113	(2.482)\\
    &	300/350	&	3.661	(0.549)	&	3.946	(0.589)	&	5.514	(1.079)	&	25.946	(2.38)	&	29.562	(3.198)\\

    \hline
    Example 2	&	50/100	&	3.685	(0.643)	&	3.842	(0.657)	&	5.174	(0.577)	&	10.815	(0.897)	&	3.465	(0.752)\\
	&	100/150	&	4.364	(0.692)	&	4.313	(0.6)	&	8.044	(0.864)	&	15.401	(0.693)	&	6.708	(1.477)\\
	&	150/200	&	4.865	(0.778)	&	5.009	(0.822)	&	10.835	(1.346)	&	17.974	(3.505)	&	9.739	(1.969)\\
	&	200/250	&	5.292	(0.912)	&	5.523	(0.77)	&	11.22	(2.208)	&	17.198	(1.18)	&	11.754  (2.444)\\
	&	250/300	&	5.648	(0.819)	&	5.601	(0.8)	&	11.741	(2.525)	&	21.952	(0.944)	&	14.064	(2.909)\\
	&	300/350	&	6.02	(1.017)	&	5.937	(0.97)	&	12.926	(1.343)	&	24.265	(1.126)	&	18.603	(4.136)\\

    \hline
    Example 3	&	50/100	&	3.545	(0.756)	& 3.687	(0.759)	&		7.633	(1.233)	&	12.148	(2.763)	&	3.43	(0.802)\\
	&	100/150	&	4.457	(0.931)	& 4.465	(0.876)	&		12.652	(1.894)	&	15.694	(1.044)	&	7.312	(1.419)\\
	&	150/200	& 4.81	(0.95)	&	4.854	(0.996)	&		13.439	(2.374)	&	19.272	(0.821)	&	9.632	(2.278)\\
	&	200/250	&	5.405	(1.016)	&	5.4 	(1.048)	&	15.187	(1.573)	&	19.8	   (0.819)	&	12.316	(2.953)\\
	&	250/300	&	5.601	(1.062)	&	5.6     (1.078)	&	21.126	(3.281)	&	23.467    (3.145)	&	15.17	(3.851)\\
	&	300/350	& 5.846	(1.156)	&	5.996	(1.019)	&		24.977	(3.736)	&	23.797	  (1.97)	&	17.683	(3.578)\\

    \hline
    Example 4	&	50/100	&	2.37	(0.347)	&	2.35	(0.375)	&	1.586	(0.194)	&	7.22	(0.828)	&	3.631	(0.689)\\
	&	100/150	&	2.739	(0.416)	&	2.604	(0.385)	&	2.216	(0.565)	&	11.135 (2.143)	&	6.235	(0.948)\\
	&	150/200	&	2.889	(0.454)	&	2.928	(0.466)	&	2.719	(0.339)	&	11.531	(1.652)	&	8.761	(1.129)\\
	&	200/250	&	3.285	(0.501)	&	3.034	(0.526)	&	4.335	(0.511)	&	17.214  	(0.858)	&	11.343	(1.523)\\
	&	250/300	&	3.374	(0.554)	&	3.13	(0.499)	&	4.56	(0.7)	&	17.899	(0.847)	&	13.634	(1.635)\\
	&	300/350	&	3.369	(0.474)	&	3.199	(0.445)	&	4.707	(0.641)	&	21.349	(2.387)	&	16	(1.897)\\

    \hline
    Example 5	&	50/100	&	8.618	(2.523)	&	7.82	(2.441)	&	19.892	(3.944)	&	21.436	(5.271)	&	7.407	(1.726)\\
	&	100/150	&	9.539	(3.62)	&	9.19	(3.328)	&	29.552(	5.825)	&	24.354	(9.468)	&	12.409	(2.854)\\
	&	150/200	&	10.446	(4.055)	&	10.127	(4.164)	&	32.976	(6.126)	&	30.37	(3.545)	&	18.704	(6.019)\\
	&	200/250	&	10.708	(3.1)	&	10.48	(3.435)	&	36.774	(11.858)&	35.2	(2.124)	&	22.434	(7.179)\\
	&	250/300	&	11.185	(4.586)	&	11.004	(2.912)	&	38.95	(10.453)&	35.042	(2.144)	&	27.187	(8.553)\\
	&	300/350	&	11.793	(4.53)	&	11.217	(3.17)	&	42.792	(7.437)	&	36.884	(3.324)	&	35.765	(14.271)\\

    \hline
    Example 6	&	50/100	&	41.305	(7.267)	&	38.194	(6.147)	&	175.42	(43.084)	&	46.041	(3.791)	&	96.404	(8.56)\\
	&	100/150	&	47.395	(11.129)&	43.339	(8.539)	&	577.228	(116.089)	&	50.525	(2.507)	&	141.277	(11.112)\\
	&	150/200	&	54.833	(11.516)&	49.191	(10.567)&	911.055	(143.733)	&	54.273	(4.283)	&	176.388	(11.166)\\
	&	200/250	&	55.214	(10.734)&	53.64	(10.378)&	965.482	(149.978)	&	58.603	(3.271)	&	216.531	(17.141)\\
	&	250/300	&	58.422	(12.195)&	57.681	(12.535)&	970.786	(151.313)	&	60.486	(3.149)&	238.397	(19.763)\\
	&	300/350	&	60.957	(16.07)	&	61.979	(14.473)&	1541.106	(122.54)&	66.983	(5.176)&	254.294	(18.688)\\

    \hline
    \end{tabular}}
\end{table}

\begin{table}[!htp]
    \centering
    \small
    \setlength\tabcolsep{6pt}
    \caption{Performance comparison measuring by MSE.}
    \label{tablemse}
    \scalebox{0.8}{
    \begin{tabular}{llccccc}
    \hline\hline
    Method&q/p &SIPBO&IPBO&MRCE&GFlasso&$l_{21}$\\
    \hline
    Example 1	&	50/100	&	74.432	(9.334)	&	75.062	(8.003)	&	91.942	(10.435)	&	75.948	(11.768)	&	98.057	(16.583)\\
	&	100/150	&	123.143	(18.321)	&	141.943	(15.23)	&	164.57	(19.536)	&	184.196	(24.818)	&	311.757	(54.405)\\
	&	150/200	&	179.173	(22.653)	&	189.649	(23.868)	&	189.201	(24.193)	&	214.496	(33.112)	&	668.037	(138.986)\\
	&	200/250	&	245.01	(37.385)	&	242.347	(28.585)	&	252.235	(32.821)	&	339.774	(38.332)	&	1057.786	(178.17)\\
	&	250/300	&	284.643	(39.896)	&	298.914	(39.816)	&	352.606	(43.691)	&	324.65	(46.269)	&	1305.599	(243.139)\\
	&	300/350	&	346.522	(48.875)	&	350.152	(47.6)	&	387.813	(52.197)	&	354.767	(50.618)	&	1426.625	(297.03)\\

    \hline
    Example 2	&	50/100	&	100.292	(12.295) &	87.87	(12.588) &	124.605	(12.721) &	104.476	(14.379) &	86.393	(11.933)\\
	&	100/150	&	160.682	(21.676)	&	185.486	(22.667)	&	243.102	(30.001)	&	192.814	(26.295)	&	187.494	(24.153)\\
	&	150/200	&	261.463	(37.406)	&	271.176	(41.117)	&	375.2	(57.209)	&	270.335	(41.173)	&	269.721	(38.974)\\
	&	200/250	&	342.614	(43.906)	&	353.843	(50.436)	&	455.22	(68.168)	&	355.162	(54.032)	&	344.638	(45.981)\\
	&	250/300	&	438.93	(59.712)	&	433.485	(67.818)	&	516.343	(69.041)	&	487.312	(58.983)	&	441.459	(56.897)\\
	&	300/350	&	499.722	(62.062)	&	501.268	(72.839)	&	545.135	(74.816)	&	588.63	(104.513)	&	510.774	(64.842)\\

    \hline
    Example 3	&	50/100	&	112.823	(17.219) &	109.358	(16.685) &	290.959	(34.502) &	131.636	(22.531) &	114.566	(14.665)\\
	&	100/150	&	150.721	(22.785)	&	148.224	(24.635)	&	258.247	(28.302)	&	185.068	(26.951)	&	154.398	(22.993)\\
	&	150/200	&	214.921	(35.788)	&	212.518	(28.321)	&	372.831	(45.319)	&	248.555	(37.556)	&	217.324	(30.919)\\
	&	200/250	&	272.351	(35.684)	&	265.968	(42.242)	&	504.421	(61.769)	&	295.426	(39.491)	&	277.453	(40.29)\\
	&	250/300	&	322.316	(43.068)	&	325.821	(38.797)	&	533.424	(57.949)	&	349.359	(50.062)	&	330.974	(45.71)\\
	&	300/350	&	383.329	(54.345)	&	378.803	(55.384)	&	635.648	(77.486)	&	398.867	(55.933)	&	391.392	(49.231)\\

    \hline
    Example 4	&	50/100	&	104.337	(13.507) &	106.326	(13.6)	& 165.849	(20.923) &	103.611	(18.381) &	94.562	(14.458)\\
	&	100/150	&	212.248	(27.486)	&	210.167	(32.455)	&	286.221	(39.322)	&	238.317	(42.759)	&	210.459	(27.697)\\
	&	150/200	&	304.52	(37.985)	&	301.367	(38.091)	&	379.841	(61.241)	&	298.087	(47.343)	&	329.268	(47.992)\\
	&	200/250	&	397.409	(62.124)	&	409.745	(57.091)	&	448.343	(64.992)	&	441.319	(69.121)	&	494.186	(77.676)\\
	&	250/300	&	482.745	(66.318)	&	512.378	(64.534)	&	542.306	(75.812)	&	606.742	(81.311)	&	617.769	(87.758)\\
	&	300/350	&	583.534	(69.262)	&	552.463	(79.183)	&	634.758	(91.204)	&	644.507	(80.777)	&	818.883	(135.6)\\

    \hline
    Example 5	&	50/100	&	70.843	(8.509)	&	64.349	(10.539) &	135.236	(22.797) &	161.106	(23.884) &	68.961	(10.522)\\
	&	100/150	&	99.264	(13.742)	&	96.601	(12.94)	&	293.224	(53.774)	&	236.591	(35.643) &	121.266	(20.185)\\
	&	150/200	&	139.949	(19.728)	&	142.642	(22.954)	&	608.118	(143.389)	&	250.965	(37.369)	&	191.828	(33.605)\\
	&	200/250	&	179.505	(25.078)	&	178.478	(35.851)	&	755.424	(152.795)	&	280.754	(46.019)	&	265.846	(46.529)\\
	&	250/300	&	213.454	(29.713)	&	233.199	(26.082)	&	787.591	(181.32)	&	312.528	(41.866)	&	298.043	(56.961)\\
	&	300/350	&	279.127	(38.611)	&	275.316	(44.881)	&	954.476	(254.364)	&	368.401	(47.717)	&	399.275	(67.858)\\

    \hline
    Example 6	&	50/100	&	227.65	(26.301) &	220.526	(28.489) &	10492.101	(1365.538) &	239.036	(32.147)	&	182.524	(27.071)\\
	&	100/150	&	189.344	(38.721)	&	178.4	(22.219)	&	14980.162	(1352.5)	&	278.353	(35.763)	&	379.67	(83.594)\\
	&	150/200	&	458.749	(63.739)	&	446.856	(47.392)	&	18712.211	(1405.634)	&	462.933	(51.438)	&	513.559	(69.026)\\
	&	200/250	&	515.489	(57.318)	&	497.617	(55.227)	&	18673.238	(1212.813)	&	518.326	(67.041)	&	645.621	(83.425)\\
	&	250/300	&	558.192	(75.13)	    &	519.523	(74.234)	&	17515.079	(1281.188)	&	530.904	(61.88)	&	710.988	(72.372)\\
	&	300/350	&	564.5	(97.436)	&	557.289	(82.574)	&	34069.434	(1663.548)	&	569.375	(91.526)	&	796.409	(106.86)\\

    \hline
    \end{tabular}}
\end{table}

\section{Analysis of Financial Datasets}
In this section, we evaluate the performance of IPBO to analyse the index
tracking problem. In economics and finance, an index is a measure to track markets or economic health. We provide here a brief description of index tracking: it is a popular passive portfolio management strategies in fund management and aims to replicate the movement of a financial index using a small set of financial assets, e.g. stocks. To reduce the transactional cost, a good passive portfolio strategy would match the performance of index as closely as possible with the asset portfolio within as few stocks as possible.
Linear regression model is widely used in the stock market, i.e. \citet{yyh2014lasso}, \citet{yyh2014elastic}, \citet{yyh2016adaptive}, \citet{fan2012vast}, \citet{Benidis2018Sparse}. The proposed method is appropriate for this problem since: i) the numbers of both responses (indices) and predictors (stocks) are large and ii) both responses and predictors have their own complex correlation structures. Reviewing results from the literature, the applications of multivariate regression model mainly focus on the biological systems while the applications in financial modeling is still relatively few. We apply the proposed method to a financial data set and aim to obtain some financially meaningful results.

The data come from Yahoo, from Jan. 2017 to May. 2019. Here and in what follows, We consider 37 rolling periods and divide each period into training (= 100 days) and testing (= 20 days) parts.
\vspace{.3 cm}

% \subsection{Analysis of Dow Jones Industrial Average}
% \subsection{Example 1}
\noindent\textbf{Example 1.}\\
In this example, we track four types of prices of Dow Jones Industrial Average, i.e. closing price, high price, low price and opening price, and use the 30 constituent stocks with their four prices as predictors. A multivariate regression model is constructed with four responses and 120 predictors, hopefully resulting in a precise estimation that each index price would find a fraction of predictors consisting of the exact price type of constituent stocks.

\begin{figure}[!htp]
    \centering
    \includegraphics[width=.48\textwidth,height=.47\columnwidth]{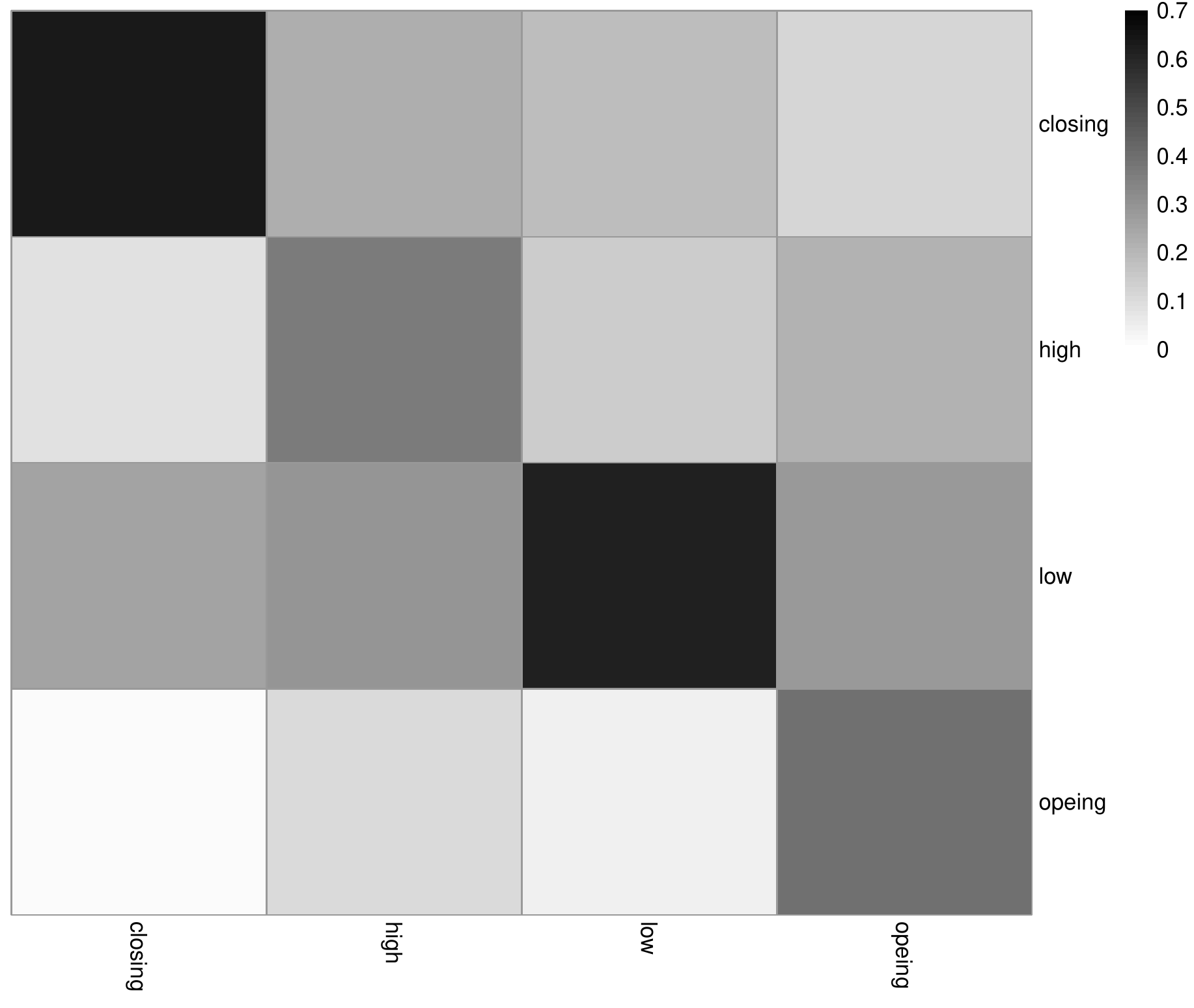}
    \caption{Columns represent four price types of responses; rows represent four price types of predictors; blocks represent the proportion of nonzero elements selected from each price type.}
    \label{figpercent}
\end{figure}

We apply the two-stage algorithm \eqref{eq beta} to solve IPBO and use 5-fold cross-validation to select the tuning parameters in the first step. In the second step, we do not use the validation or cross-validation approach to select the tuning parameters; instead, we choose the tuning parameters such that the number of the selected predictors is 10. Results are shown in Figure~\ref{figpercent}, where the columns and rows represent the four price types of index and that of constituent stocks respectively. The color denotes the average degree over 37 periods that the proportion of nonzero elements selected from each price type. As shown in Figure~\ref{figpercent}, every diagonal element has the largest proportion, which means for each response its fraction of predictors mostly comes from the matched price type.
\vspace{.3 cm}

\begin{figure}[!htp]
    \centering
    \includegraphics[width=.48\textwidth,height=.46\columnwidth]{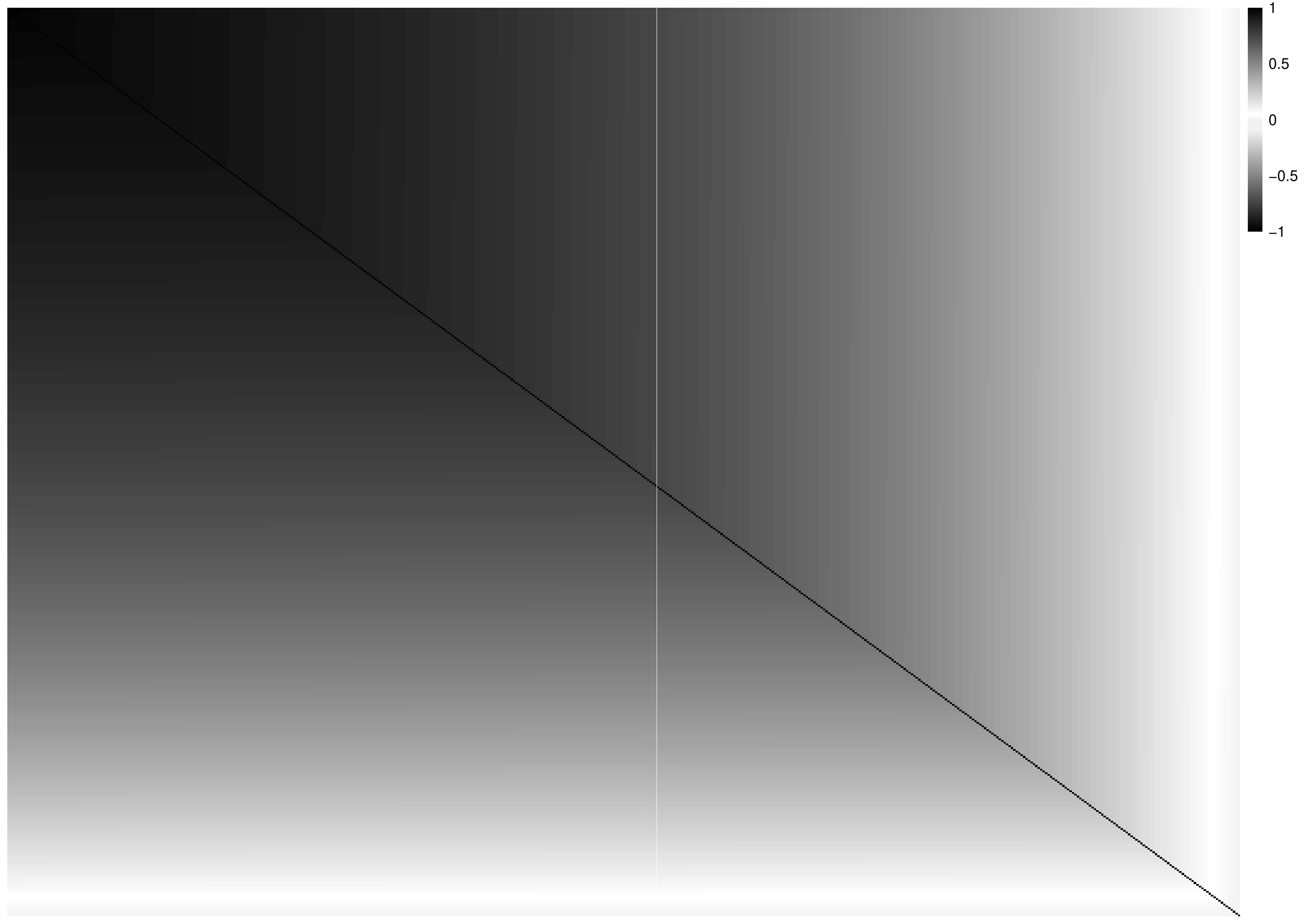}
	\caption{Estimated covariance of all the constituent stocks of S$\&$P500,NDX, DJI and XMI.}
    \label{figcor}
\end{figure}

% \subsection{Analysis of four indices tracking}
% \subsection{Example 2}
\noindent\textbf{Example 2.}\\
In this example, we track four indices, i.e. S$\&$P500, NASDAQ-100 (NDX), Dow Jones Industrial Average (DJI) and NYSE Arca Major Market Index (XMI). All of them are representative indices of the American stock market. And the constituent stocks of four indices are partially overlapped, leading highly correlations within responses.
Let $y_{kt}$ represents the closing price of four indices and $x_{jt}$ represents the closing price of the $j$th constituent stock. We describe the relationship between $x_{jt}$ and $y_{kt}$ by the multivariate regression model that for $ k=1,\dots,4 $, \[ y_{kt} = \sum^{523}_{j=1}\beta_{jk}x_{jt} + e_{kt}. \]
Figure~\ref{figcor} and Table~\ref{tableindex} show the estimated covariance of all the constituent stocks and the sample correlations of the indices respectively.

\begin{figure}[!htp]
    \centering
    \subfigure[NASDAQ-100]
    {\includegraphics[width=.48\textwidth,height=.25\columnwidth]{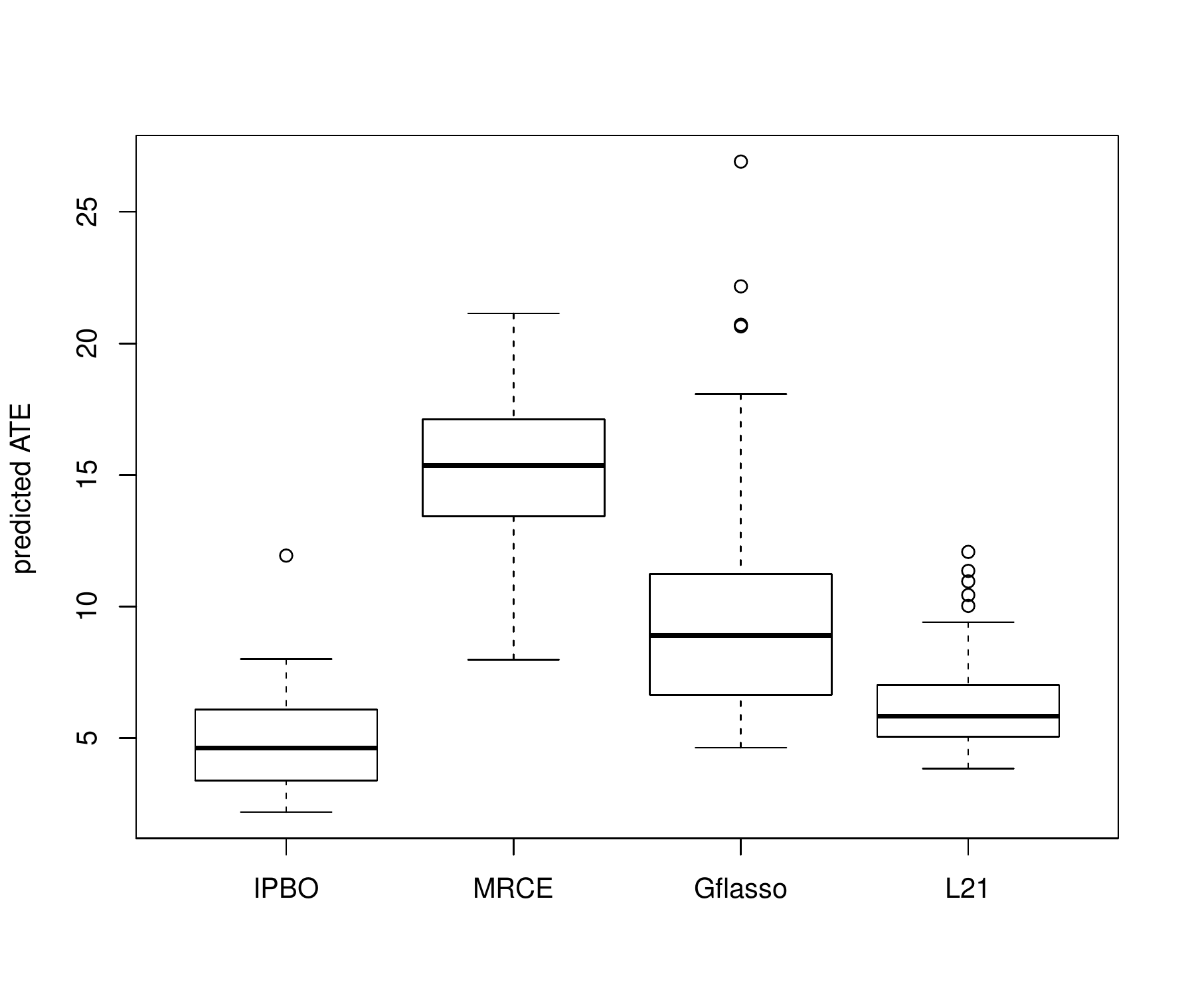}}
    \subfigure[NYSE Arca Major Market index]
    {\includegraphics[width=.48\textwidth,height=.25\columnwidth]{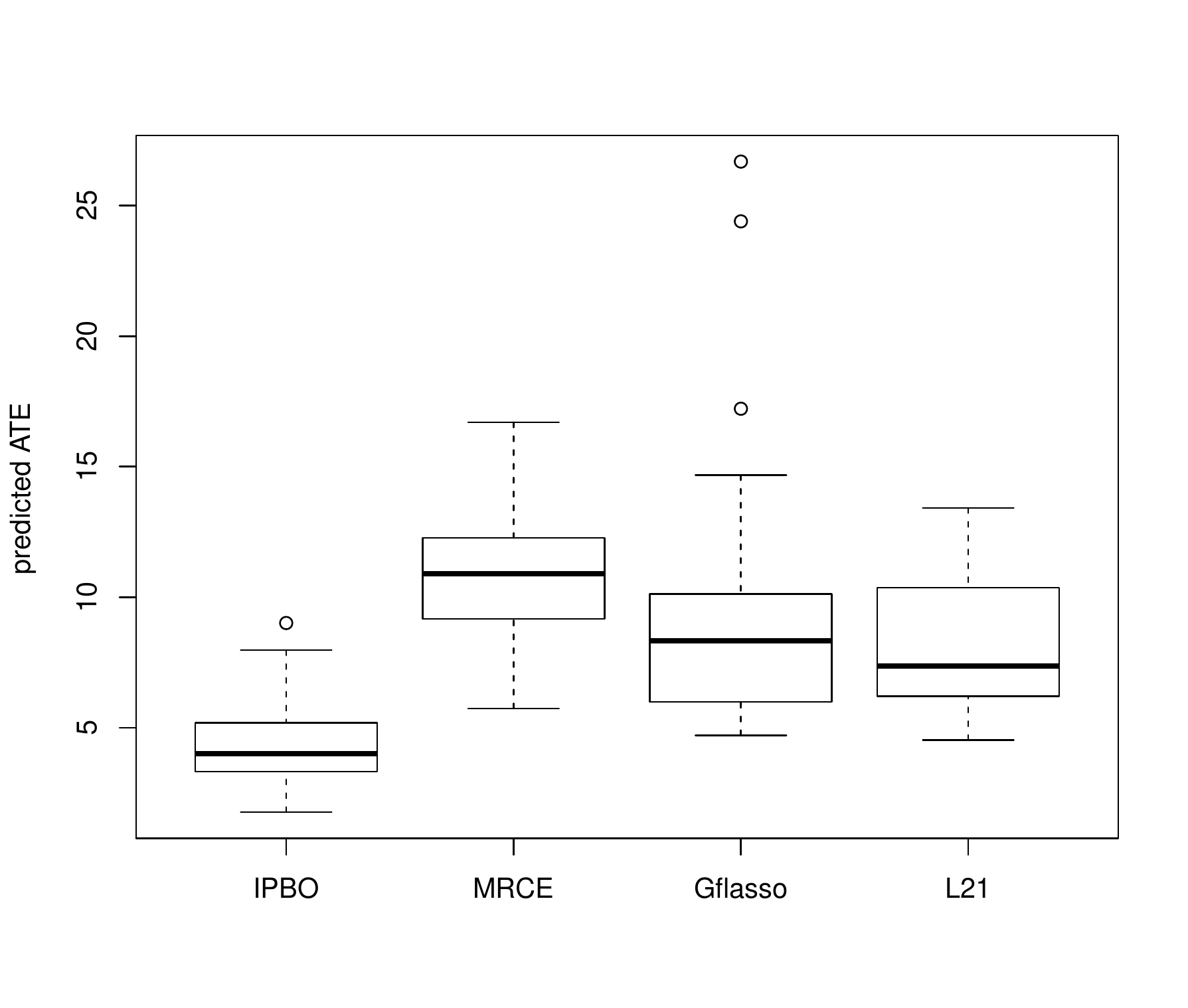}}
    \subfigure[Dow Jones Industrial Average]
    {\includegraphics[width=.48\textwidth,height=.25\columnwidth]{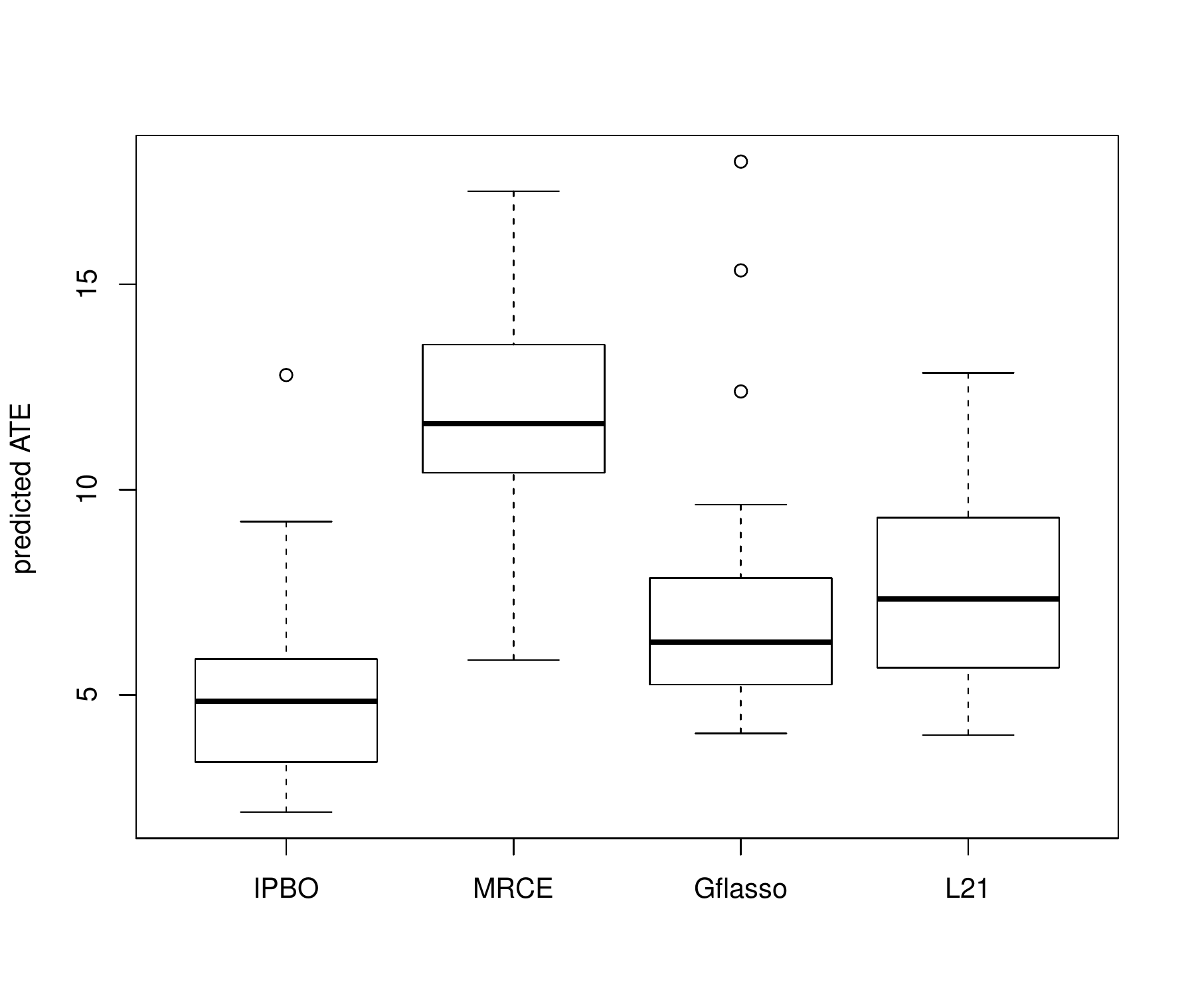}}
    \subfigure[S$\&$P500]
    {\includegraphics[width=.48\textwidth,height=.25\columnwidth]{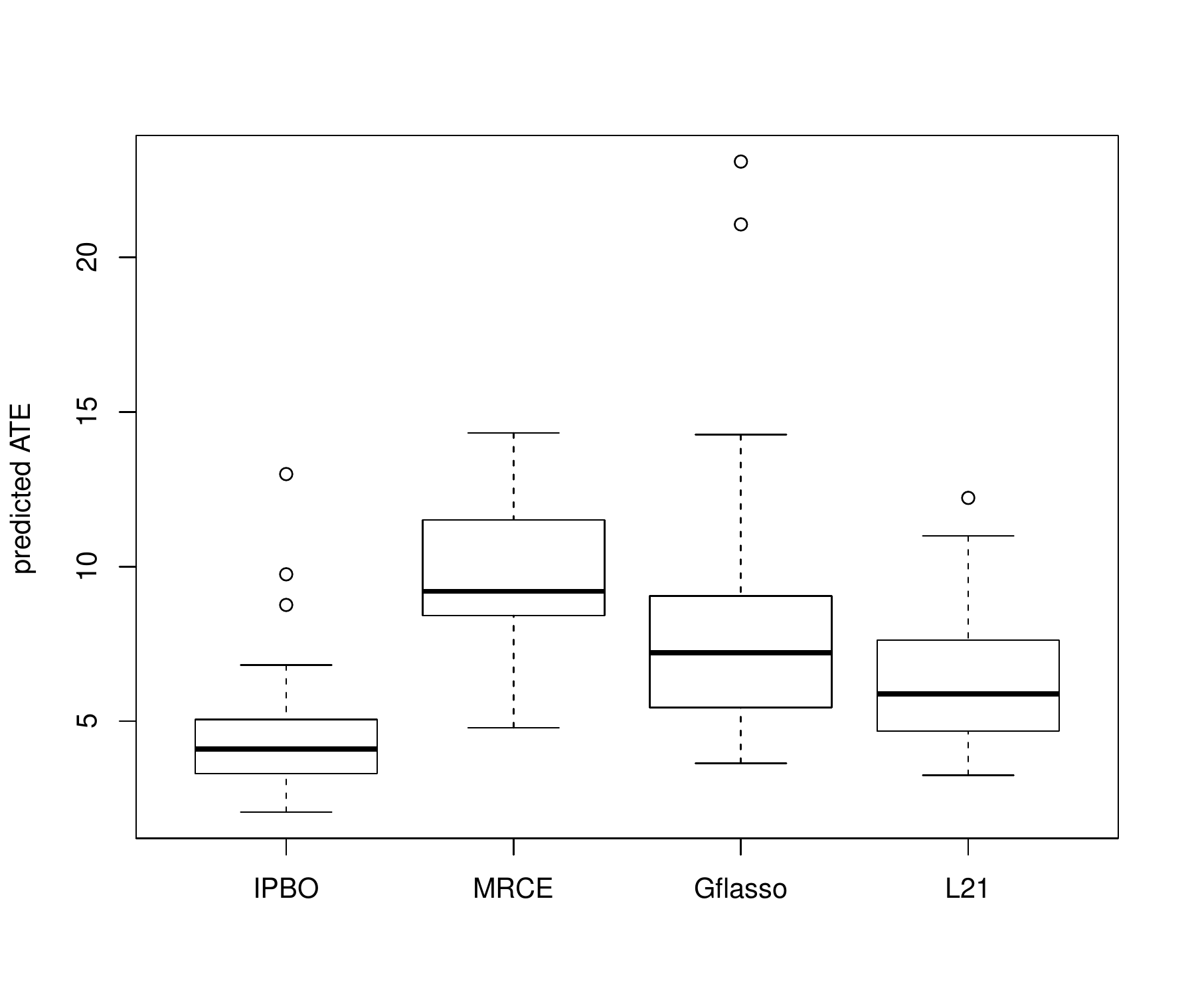}}
	\caption{Performance comparison of predicted ATE(\%).}
    \label{figate}
\end{figure}

\begin{figure}[!htp]
    \centering
    \subfigure[NASDAQ-100]
    {\includegraphics[width=.48\textwidth,height=.25\columnwidth]{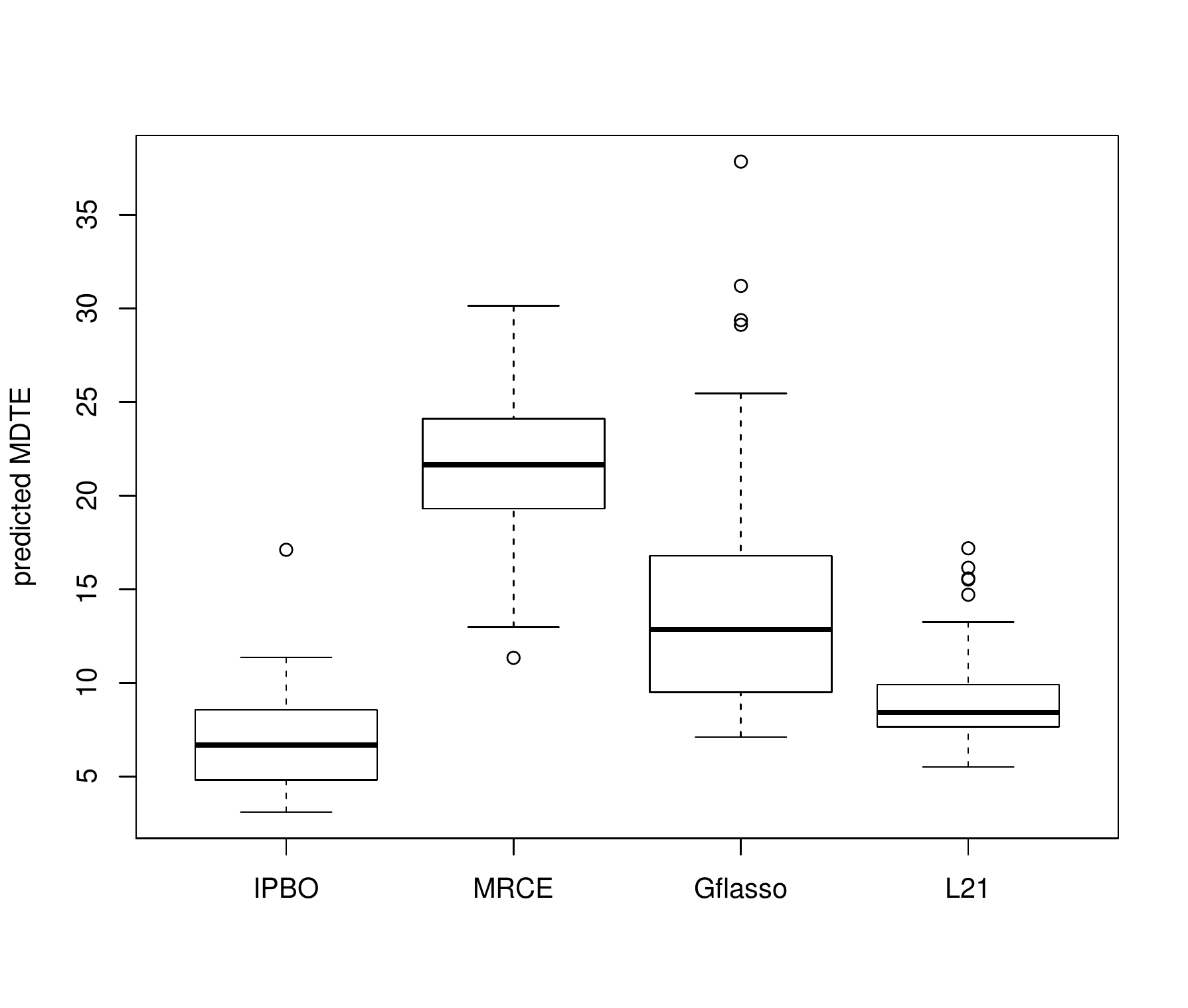}}
    \subfigure[NYSE Arca Major Market index]
    {\includegraphics[width=.48\textwidth,height=.25\columnwidth]{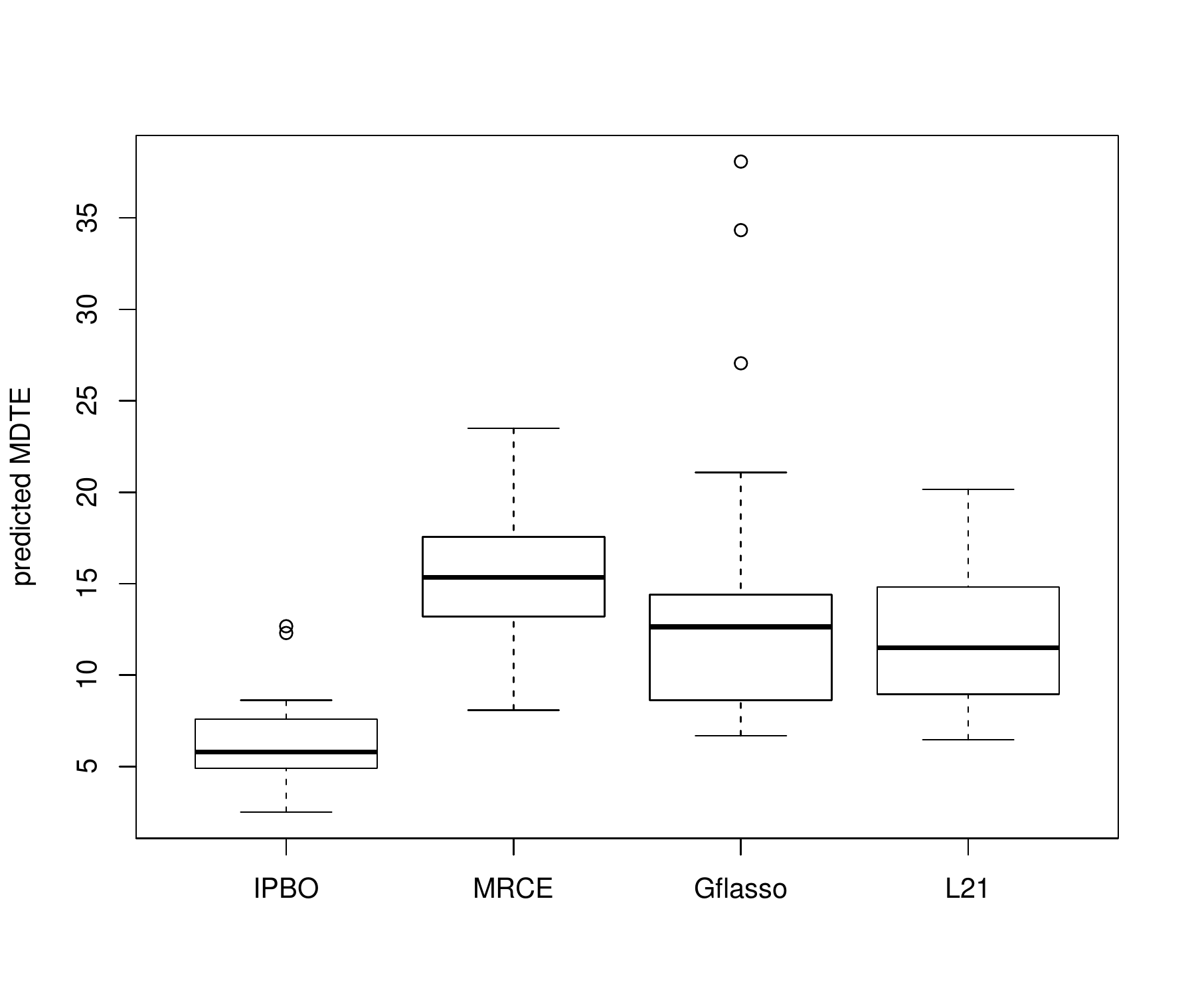}}
    \subfigure[Dow Jones Industrial Average]
    {\includegraphics[width=.48\textwidth,height=.25\columnwidth]{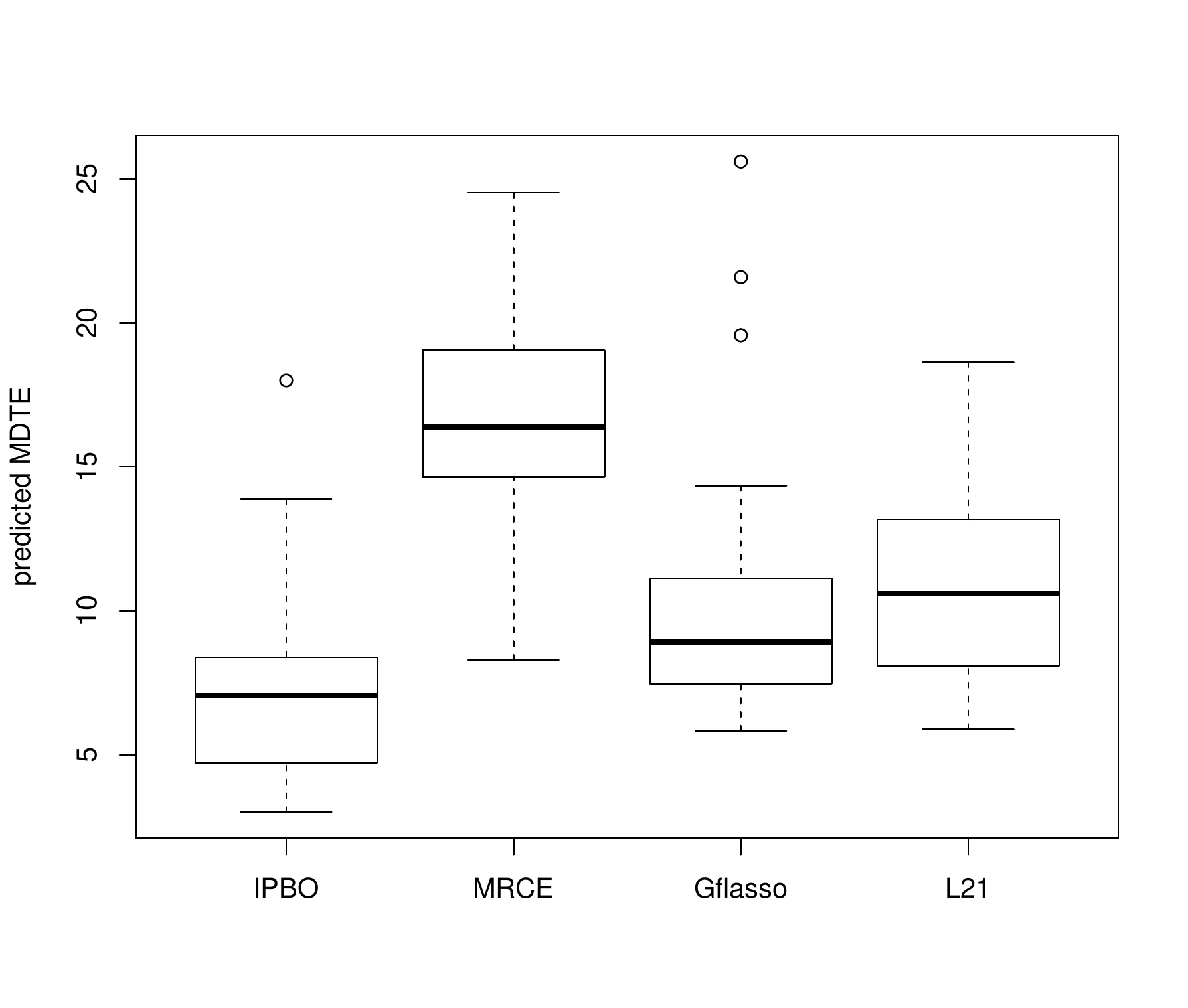}}
    \subfigure[S$\&$P500]
    {\includegraphics[width=.48\textwidth,height=.25\columnwidth]{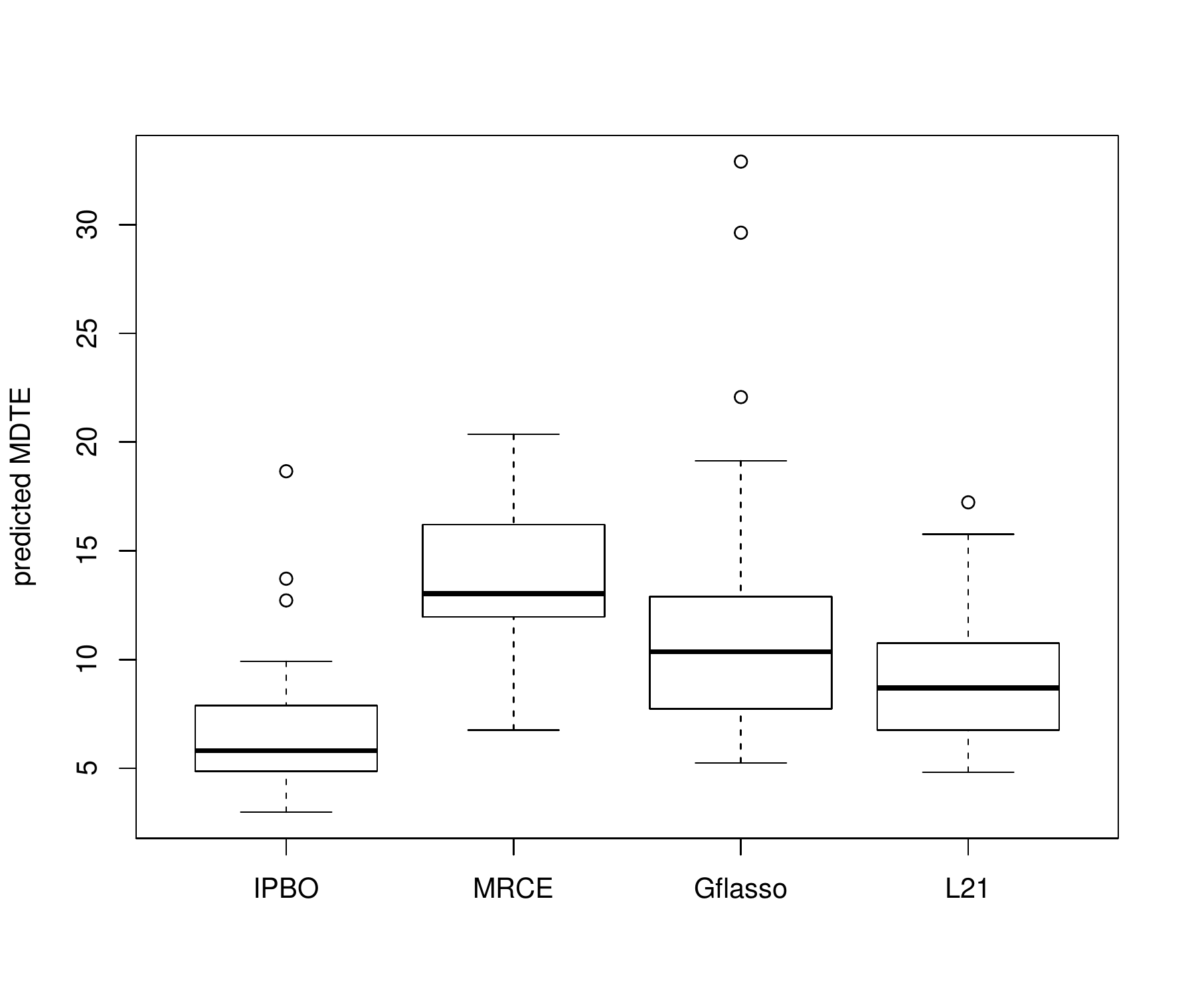}}
    \caption{Performance comparison of predicted MDTE(\textpertenthousand).}
    \label{figmdte}
\end{figure}

\begin{figure}[!htp]
    \centering
    \subfigure[NASDAQ-100]
    {\includegraphics[width=.8\textwidth,height=.25\columnwidth]{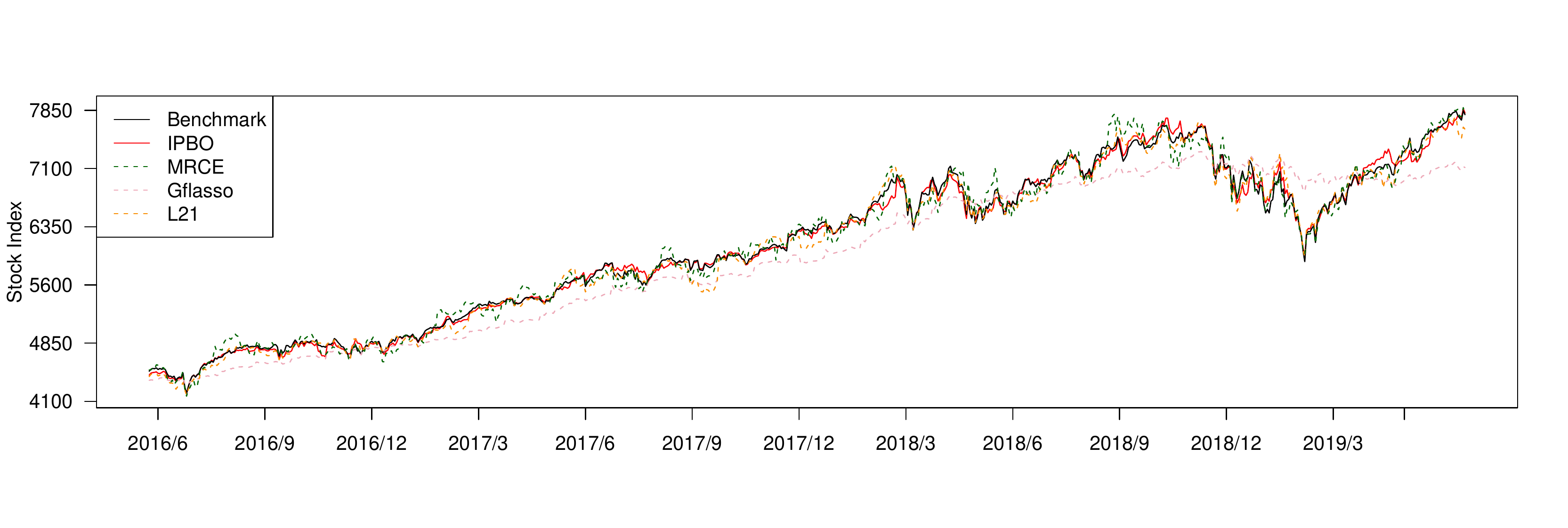}}
    \subfigure[NYSE Arca Major Market]
    {\includegraphics[width=.8\textwidth,height=.25\columnwidth]{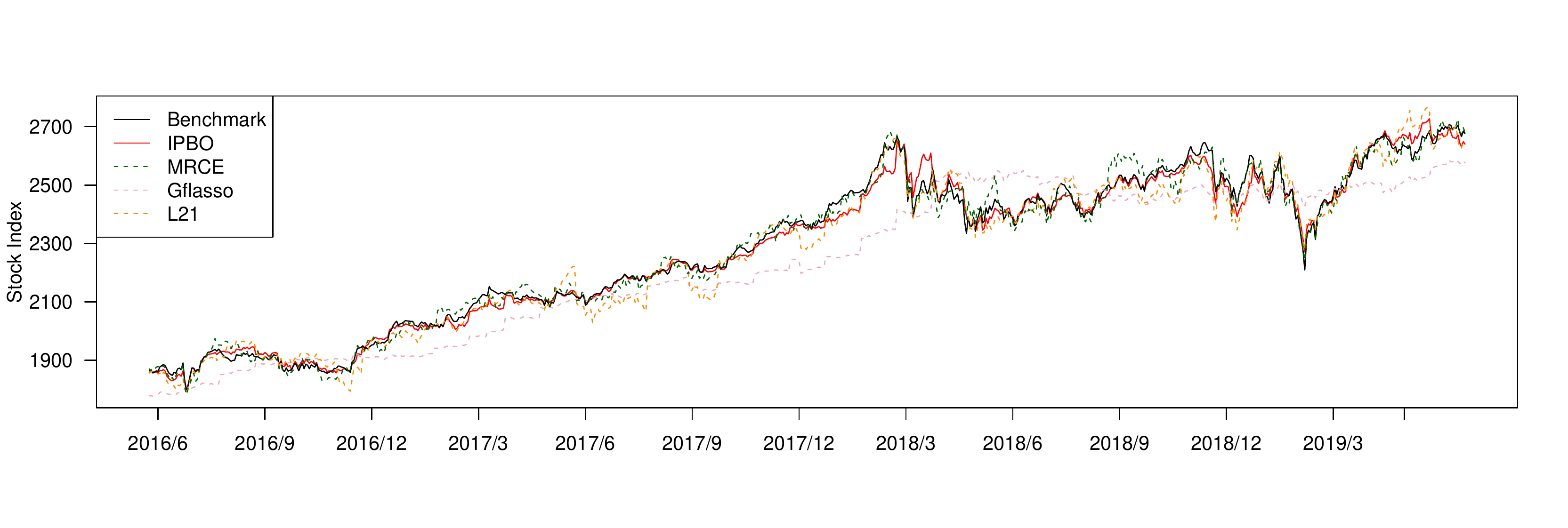}}
    \subfigure[Dow Jones Industrial Average]
    {\includegraphics[width=.8\textwidth,height=.25\columnwidth]{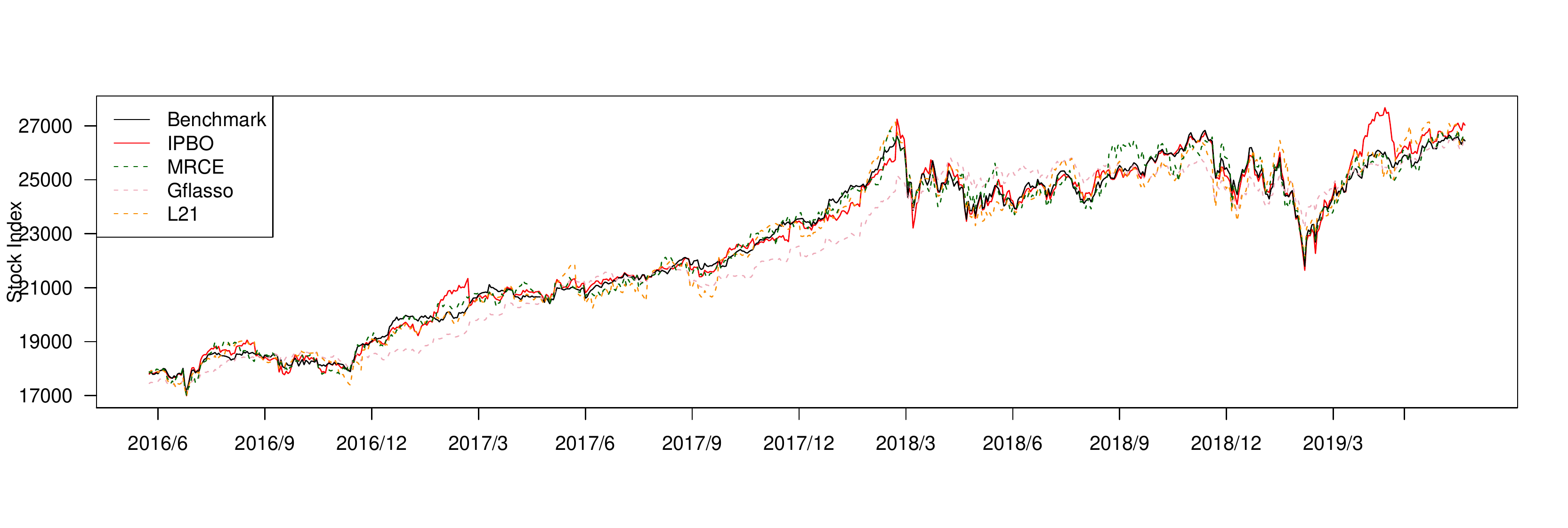}}
    \subfigure[S$\&$P500]
    {\includegraphics[width=.8\textwidth,height=.25\columnwidth]{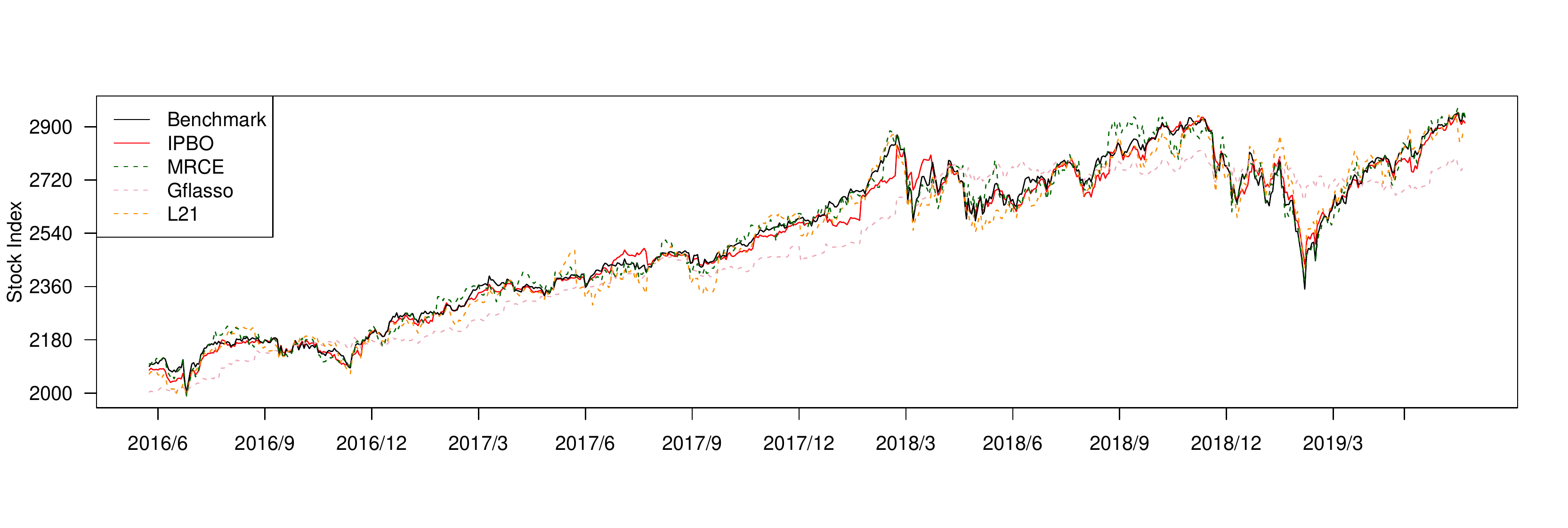}}
	\caption{Performance comparison of index tracking.}
    \label{figindex}
\end{figure}

We use the Annual Tracking Error (ATE) and the Magnitude of the Daily Tracking Error (MDTE) to be the measurements. Both are standard measures used in the financial industry to assess the performance of tracking. Set daily return rate $r_t=(y_t-y_{t-1})/y_{t-1}$ and $\textmd{error}_t=r_t-\hat r_t$. We have  \[\textmd{ATE} =\sqrt{252} \sqrt {\dfrac{\sum (\textmd{error}_t - \textmd{mean(error)})^2}{T - 1}}\]
and
\[\textmd{MDTE}=\dfrac{\sum \textmd{error}_t^2}{T - 1}.\]
Four methods are compared, i.e. IPBO, MRCE, Gflasso and $ l_{21} $. We do not use the validation or cross-validation approach to select the tuning parameter; instead, we choose the tuning parameter for each method such that the number of selected stocks is 40. The forecasting results are presented in Figure~\ref{figate}, Figure~\ref{figmdte} and Figure~\ref{figindex}. As shown in Figure~\ref{figate} and Figure~\ref{figmdte},
IPBO uniformly outperform other methods in both mean values and deviations, i.e. its predicted ATE are nearly between $3\%$-$5\%$ and its predicted MDTE are 6\textpertenthousand-8\textpertenthousand~while other methods are greater than $6\%$ and 9\textpertenthousand. In Figure~\ref{figindex}, IPBO is closest to the benchmark and other methods have large fluctuations.

\begin{table}[!htp]
    \centering
    \small
    \setlength\tabcolsep{6pt}
    \caption{Sample Correlations of four indices.}
    \label{tableindex}
    \begin{tabular}{ccccc}
    \hline\hline
    &XMI & NDX & DJI & S$\&$P500\\
    \hline
    XMI & 1 & 0.978 & 0.997 & 0.989\\
    NDX & 0.978 & 1 & 0.985 & 0.99\\
    DJI & 0.997 & 0.985 & 1 & 0.994\\
    S$\&$P500 & 0.989 & 0.99 & 0.994 & 1\\
    \hline
    \end{tabular}
\end{table}

\section{Summary}
We study the multivariate regression models and propose an efficient method called Interaction Pursuit Biconvex Optimization (IPBO). We assume that both predictors and responses are derived from different multivariate normal distributions with general covariance matrices, while correlation structures within are always complex and interact on each other based on the regression function. The proposed method uses the Laplacian quadratic associated with the graph information to improve estimation efficiency by promoting smoothness among the coefficients between the linked variables. We compare this method with several existing methods, showing that capturing the graph information from responses instead of from noises has advantages on both estimation and computational efficiency.

This paper focuses on building the structured sparsity penalty to encourages the shared structure between the network and the regression coefficients. It would be interesting to extend the structured sparsity penalty to other forms of adjacency measure which have been successfully used in network analysis \citep{huang2011laplacian}. It would be also interesting to extend the setting to more complicated settings, such as heavy-tailed noise, influential observations.

\section*{Acknowledgement}
This work was partially supported by the National Natural Science Foundation of China (Grant No. 11671059), the Fundamental Research Funds for the Central Universities (Grant No. QL18010) and the Program for Innovation Research in Central University of Finance and Economics.

\bibliographystyle{apalike}
\bibliography{C:/reference/reference}
\end{document}